\documentclass[doublespacing]{elsart}

\usepackage{times}
\usepackage{graphicx}
\usepackage{natbib}

\begin{document}

\begin{frontmatter}

\title{Modification of the Pattern Informatics Method for Forecasting Large
       Earthquake Events Using Complex Eigenvectors}

\author[addr1,addr2]{J. R. Holliday} \ead{holliday@cse.ucdavis.edu}
\author[addr1,addr2]{J. B. Rundle} \ead{jbrundle@ucdavis.edu}
\author[addr3]{K. F. Tiampo} \ead{ktiampo@uwo.ca}
\author[addr4]{W. Klein} \ead{klein@buphyc.bu.edu}
\author[addr5]{A. Donnellan} \ead{donnellan@jpl.nasa.gov}

\address[addr1]{
  Center for Computational Science and Engineering,
  University of California,
  Davis, California, USA.
}

\address[addr2]{
  Department of Physics,
  University of California,
  Davis, California, USA.
}

\address[addr3]{
  Department of Earth Sciences,
  University of Western Ontario,
  London, Ontario, CANADA.
}

\address[addr4]{
  Department of Physics,
  Boston University,
  Boston, Massachusetts, USA.
}

\address[addr5]{
  Earth and Space Sciences Division,
  Jet Propulsion Laboratory,
  Pasadena, California, USA.
}

\begin{abstract}
Recent studies have shown that real-valued principal component
analysis can be applied to earthquake fault systems for forecasting
and prediction.  In addition, theoretical analysis indicates that
earthquake stresses may obey a wave-like equation, having solutions
with inverse frequencies for a given fault similar to those that
characterize the time intervals between the largest events on the
fault.  It is therefore desirable to apply complex principal component
analysis to develop earthquake forecast algorithms.  In this paper we
modify the Pattern Informatics method of earthquake forecasting to
take advantage of the wave-like properties of seismic stresses and
utilize the Hilbert transform to create complex eigenvectors out of
measured time series.  We show that Pattern Informatics analyses using
complex eigenvectors create short-term forecast hot-spot maps that
differ from hot-spot maps created using only real-valued data and
suggest methods of analyzing the differences and calculating the
information gain.
\end{abstract}


\end{frontmatter}


\section{Introduction}
Principal component analysis (PCA) is a mathematical procedure that
transforms a set of correlated variables into a smaller set of
uncorrelated variables called principal components. The first
principal component accounts for as much of the variability in the
data as possible, and each succeeding component attempts to account
for the remaining variability.  \cite{Savage88} introduced PCA to the
seismic community by using it to decompose time series data into a
complete set of orthonormal subspaces that isolate spatial and
temporal eigensources.

Complex principal component analysis is an extension of classical
principal component analysis in which the spatial basis vectors
represent the eigenfunctions of a complex correlation matrix. It is
closely related to principal oscillation pattern (POP) analysis, in
which the oscillating basis pattern states are the eigenfunctions of a
deterministic feedback matrix \citep{Penland89} (both techniques
empirically identify time-dependent spatial patterns in a multivariate
time series of geophysical or other data).  POP analysis has been
shown to be reasonably successful in forecasting El Ni\~no-Southern
Oscillation (ENSO) events up to a year in advance \citep{WuAD94}. In
complex PCA, a real-valued time series is analytically continued into
the complex-valued domain by means of a Hilbert transform
\citep{Horel84}, then the $N \times N$ complex correlation matrix is
formed via cross-correlation of the $N$ independent time series. These
methods have been applied extensively in the atmospheric and ocean
sciences \citep{Penland89, Burger93, ZhangDW97, Egger99, KimN99}.

The primary benefit of complex PCA compared to other analysis
procedures is that it allows propagating features within the time
series to be detected and dissected in terms of their spatial and
temporal behavior \citep{Horel84}.  In particular, localized
propagating phenomena, if they exist, can be easily detected.
Classical PCA, for example, allows only the detection of standing
oscillations.

Recently it has been shown that the same kinds of real-valued PCA
analysis can be applied to earthquake fault systems for forecasting
and prediction \citep{RundleKTG00, TiampoRMG02, TiampoRMK02}. It is
known that earthquakes recur in complex cycles, similar to ENSO
events, albeit with the larger earthquake events having substantially
longer time scales \citep{Scholz90} than those that apply to
ENSO--typically a decade or less. In addition, theoretical analysis
\citep{Klein04} indicates that earthquake stresses may obey a
wave-like equation, having solutions with inverse frequencies for a
given fault similar to those that characterize the time intervals
between the largest events on the fault. It is of considerable
interest to apply complex PCA and POP analysis to develop earthquake
forecast algorithms, taking account of the complex cyclic and
quasi-periodic nature of these events.

A problem with this approach is that earthquake event time series are
typically not continuous and differentiable, but instead are point
processes, both in space and in time. In addition, high quality
measurements of earthquakes have only been comprehensively observed
with instruments for a few decades, so the complete (high-density)
time series that are available are relatively short compared to the
recurrence periods for large earthquakes of hundreds of years and
longer.  The Pattern Informatics (PI) method for earthquake
forecasting is well suited for these types of impulsive time series
and performs very well with data sets much shorter than the recurrence
periods for large earthquakes events
\citep{HollidayRTKD05}.  As such, it is an ideal candidate for modification
to use complex eigenfunctions and eigenvectors.  Assuming that seismic
phenomena are analytic, causality considerations allow us to apply the
Cauchy Riemann dispersion relations \citep{ArfkenW01} and analytically
continue the measured time series from the real axis into the entire
upper half-plane of complex space.  In this new space we propose to
utilize the PI method.


\section{Modified Method}
Our modified PI method is based on the idea that the future time
evolution of seismicity can be described by pure phase dynamics
\citep{MoriK98, RundleKGT00, RundleKTG00}, hence a complex seismic
phase function $\hat{\mathcal S}({\bf x}_i, t_b, t)$ is constructed
and allowed to rotate in its Hilbert space.  This modified
representation of the input data serves two purposes.  First, a
complex Hilbert space allows detection both of standing oscillations
and traveling waves \citep{Horel84}.  This is important for
identifying the quasi-periodic nature of seismicity.  Second, the
construction allows for interference between the real and imaginary
parts of the phase function.  This interference helps correlate
geographic locations which are spatially separated.

To create our phase function, the geographic area of interest is
partitioned into $N$ square bins centered on a point ${\bf x}_i$ and
with an edge length $\delta x$ determined by the nature of the
physical system.  For our analysis we chose $\delta x = 0.1^\circ
\approx 11$km, corresponding to the linear size of a magnitude $M \sim
6$ earthquake.  Within each box, a time series $\psi_{obs}({\bf x}_i,
t)$ is defined by counting how many earthquakes with magnitude greater
than $M_{min}$ occurred during the time period $t$ to $t + \delta t$.
These time series are interpreted as the real-valued portion of an
analytic signal, and thus the entire signal is recreated by combining
$\psi_{obs}$ with its Hilbert transform:
\begin{equation}
\psi_{obs}({\bf x}_i,t) \to \Psi({\bf x}_i,t) \equiv
\psi_{obs} + \tilde{\psi}_{obs},
\end{equation}
where $\tilde{\psi}_{obs}({\bf x}_i,t) = \frac{1}{\pi} {\mathcal
P}\int_{-\infty}^{\infty} \frac{\psi({\bf x}_i,\tau) d\tau}{t-\tau}$
and Cauchy principal value integration is specified
\citep{Bracewell99}.  Next, the activity rate function ${\mathcal
S}({\bf x}_i, t_b, T)$ is defined as the average rate of occurrence of
earthquakes in box $i$ over the period $t_b$ to $T$:
\begin{equation}
{\mathcal S}({\bf x}_i, t_b, T) =
\frac{\sum_{t = t_b}^T \Psi({\bf x}_i, t)}{T - t_b}
.
\end{equation}
If $t_b$ is held to be a fixed time, ${\mathcal S}({\bf x}_i, t_b, T)$
can be interpreted as the $i$th component of a general, time-dependent
vector evolving in an $N$-dimensional space \citep{TiampoRMK02}.
Furthermore, it can be shown that this $N$-dimensional correlation
space is defined by the eigenvectors of an $N \times N$ correlation
matrix \citep{RundleKGT00, RundleKTG00}.  In order to remove the final
free parameter in the system--the choice of base year--changes in the
activity rate function are then averaged over all possible base-time
periods:
\begin{equation}
{\underline{\mathcal S}}({\bf x}_i, t_0, T) =
\frac{\sum_{t_b = t_0}^{T} {\mathcal S}({\bf x}_i, t_b, T)}
{T - t_0}
.
\end{equation}
The base averaged activity rate function is then normalized by
subtracting the spatial mean over all boxes and scaling to give a
unit-norm:
\begin{equation}
\hat{\underline{\mathcal S}}({\bf x}_i, t_0, T) =
\frac{{\underline{\mathcal S}}({\bf x}_i, t_0, T) -
\frac{1}{N} \sum_{j = 1}^N {\underline{\mathcal S}}({\bf x}_j, t_0, T)}
{\sqrt{\sum_{j = 1}^N [{\underline{\mathcal S}}({\bf x}_j, t_0, T) -
\frac{1}{N} \sum_{k = 1}^N {\underline{\mathcal S}}({\bf x}_k, t_0, T)]^2}}
.
\end{equation}
The requirement that the rate functions have a constant norm helps
remove random fluctuations from the system.  Following the assumption
of pure phase dynamics \citep{RundleKGT00, RundleKTG00}, the important
changes in seismicity will be given by the change in the normalized
base averaged activity rate function from the time period $t_1$ to $t_2$:
\begin{equation}
\Delta \hat{\underline{\mathcal S}}({\bf x}_i, t_0, t_1, t_2) =
\hat{\underline{\mathcal S}}({\bf x}_i, t_0, t_2) -
\hat{\underline{\mathcal S}}({\bf x}_i, t_0, t_1)
.
\end{equation}
This is simply a pure rotation of the $N$-dimensional unit vector
$\hat{\underline{\mathcal S}}({\bf x}_i, t_0, T)$ through time.  
Finally, the probability of change of activity in a given box is
deduced from the square of its base averaged, mean normalized change
in activity rate:
\begin{equation}
P({\bf x}_i, t_0, t_1, t_2) =
[\Delta \hat{\underline{\mathcal S}}({\bf x}_i, t_0, t_1, t_2)]^{\star} \times
[\Delta \hat{\underline{\mathcal S}}({\bf x}_i, t_0, t_1, t_2)]
,
\end{equation}
where multiplication and complex conjugation are indicated.  In phase
dynamical systems, probabilities are always related to the square of
the associated vector phase function \citep{MoriK98, RundleKTG00}.
This probability function is often given relative to the background by
subtracting off its spatial mean:
\begin{equation}
P'({\bf x}_i, t_0, t_1, t_2) \Rightarrow P({\bf x}_i, t_0, t_1, t_2) - \mu
,
\end{equation}
Where $\mu = \frac{1}{N}\sum_{j=1}^N P({\bf x}_j, t_0, t_1, t_2)$ and
$P'$ indicates the probability of change in activity is measured
relative to the background.


\section{Application Of The Method}
\begin{figure*}
\noindent\includegraphics[width=\textwidth]{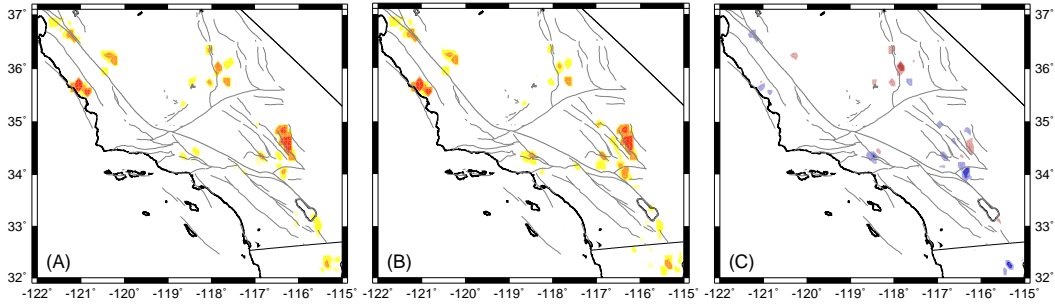}
\caption{Logarithmic seismic hot-spot map for large earthquake events
with $M\ge5$ for the forecasted time period 1 August 2004 to 31 July
2009 using (A) complex eigenvectors and (B) real-valued eigenvectors.
Data from the SCEDC catalog was used below $35^\circ$ North latitude
while data from the NCEDC catalog was used above $35^\circ$ North
latitude.  Figure (C) is a difference map plotted with a linear
scale.}
\label{fig:forecast}
\end{figure*}

As an application of the modified PI method, we created a short-term
forecast seismic hot-spot map for Southern California over the time
period 1 August 2004 to 31 July 2009.  The result is shown in
Figure~\ref{fig:forecast}A.  Also presented in
Figure~\ref{fig:forecast} is the same forecast map created with
real-valued eigenvectors (\ref{fig:forecast}B) and a difference map
between the two methods (\ref{fig:forecast}C).

Two data sets were employed in this analysis, the first being the
entire historic seismic catalog from 1 January 1932 through 31 July
2004, obtained from the Southern California Earthquake Data Center
(SCEDC) on-line searchable database%
\footnote{http://www.data.scec.org/catalog\_search/index.html},
with all non-local and blast events specifically removed.  The
relevant data consists of location, in East longitude and North
latitude, and the date the event occurred.  Seismic events between
$-122^\circ$ and $-115^\circ$ longitude and between $32^\circ$ and
$35^\circ$ latitude (any depth and quality) and with magnitude greater
than or equal to $M_{min} = 3.0$ were selected.  Data from the time
period 1977-1980 is currently missing from the database but can be
found at the older Southern California Seismic Network (SCSN)
archives\footnote{http://www.data.scec.org/ftp/catalogs/SCSN/}.

The second source of data employed in this analysis was acquired from
the Northern California Earthquake Data Center (NCEDC) on-line
searchable database%
\footnote{http://quake.geo.berkeley.edu/ncedc/catalog-search.html},
with all non-local and blast events again specifically removed.  When
incorporating this catalog, seismic events between $-122^\circ$ and
$-115^\circ$ longitude and between $35^\circ$ and $37^\circ$ latitude
(any depth and quality) and with magnitude greater than or equal to
$M_{min} = 3.0$ were selected.  The necessity for utilizing this
additional catalog in our analysis arises from various earthquake
events in the vicinity of $35^\circ$ North latitude missing from the
SCEDC/SCSN catalog but present in the NCEDC collection.

The necessity of combining catalogs arises from the
fact that the SCEDC catalog is not complete in its network coverage
above the joining mark.  Most notably, it does not contain
earthquakes from the San Simeon region (location of the $M = 6.5$,
2003 event).

As can be seen in Figure~\ref{fig:forecast}, the map created using
complex eigenvectors is similar to the map created using real-valued
eigenvectors.  Important differences, however, are present.  Most
prominent are the increased emphasis of forecasted activity
surrounding $36^\circ$ North latitude, $-117.9^\circ$ East longitude
and the decreased emphasis of forecasted activity southwest of the
1999 Hector Mine events.  While future monitoring of these areas will
be necessary to help determine the accuracy and reliability of complex
PI analysis, certain measurements can be performed to estimate the
information gain.

\subsection{Entropy}
Using methods from information theory \citep{CoverT91}, we can
calculate the entropy, $H$, of our two hot-spot maps.  Entropy can be
considered a measure of disorder ({\it e.g.\/} randomness) or
``surprise'', hence maps with lower entropy contain more useful
information than maps with higher entropy.  We define entropy as
\begin{equation}
H(z) = - \sum_{i=1}^N p({\bf x}_i;z) \log p({\bf x}_i;z)
,
\end{equation}
where
\begin{equation}
p({\bf x}_i;z) = \left\{ 
{P({\bf x}_i, t_0, t_1, t_2) \atop 0} \qquad
{P({\bf x}_i, t_0, t_1, t_2) \ge z \atop P({\bf x}_i, t_0, t_1, t_2) < z}
\right.
,
\end{equation}
and the probabilities are scaled such that $\sum_{i=1}^N p({\bf x}_i)
= 1$.  This definitions allows a measurement of entropy as a function
of some lower threshold.

Performing this calculation on the two maps indicates that the complex
PI analysis does indeed yield more useful information (lower
$H$-value) than the original analysis, but only when the lower
threshold is non-zero.  With complex PCA calculations, sudden
transitions and noisy spikes are emphasized \citep{Horel84}.  Since
seismic time series data can be approximated by chains of delta
functions, we expect that calculations in the complex domain would
contain more low-level noise.  A small, non-zero threshold allows us
to measure the entropy above and relative to this noise.

\subsection{ROC Analysis}
A second measure for the accuracy of the hot-spot maps can be inferred
from relative operating characteristic (ROC) diagrams.  ROC curves are
essentially signal detection curves for binary forecasts obtained by
plotting the hit rate (y-axis) against the false alarm rate (x-axis)
over a range of different thresholds \citep{JolliffeS03}.  Originally
established for verifying tornado forecasts \citep{MurphyW87}, ROC
frameworks have recently become popular in the seismic community as
well \citep{Molchan97}.

While only one year has passed since the onset of the hot-spot
forecasts given in Figure~\ref{fig:forecast}, we can create ROC
diagrams for the two maps by considering a ``hit'' to be any box $i$
with $P({\bf x}_i, t_0, t_1, t_2) \ge z$, for some threshold $z$, that
contains a future large earthquake.  Similarly we consider a ``false
alarm'' to be any box $j$ with $P({\bf x}_j, t_0, t_1, t_2) \ge z$,
for some threshold $z$, that does not contain a future large
earthquake.  Since a successful forecast will maximize the hit rate
while minimizing the false alarm rate, a measure of the forecast
accuracy is given by the area, $A_{ROC}$ under the ROC curve.  It can
be shown that $A_{ROC} \to 1$ for a perfect forecast and $A_{ROC} \to
0.5$ for a forecast consisting of randomly distributed alarms.

Performing this calculation on the two maps again indicates that
the complex PI analysis is better correlated with future large events
(higher $A_{ROC}$-value) than the original analysis.  It is important
to consider, however, that this analysis is only using one year of
observed future seismicity.  A full analysis should be performed at the
end of the forecast interval.


\section{Conclusion}
Complex PCA is a useful tool and is ideally suited for many
applications.  There are, however, situations where the results of
complex PCA are difficult to interpret such as when both amplitude and
phase relationships must be considered.  For these types of systems,
the existence of phase information by itself suggests the need for an
analysis in the full complex domain.

The theoretical evidence that earthquake stress fields are wave-like
in nature indicates that seismicity is better studied using complex
time series.  Due to its ability to create seismic hot-spot forecast
maps using relatively short time series data and its handling of
impulsive data sets, the PI method is naturally extended to this
complex domain.

In our five year seismic hot-spot forecast for southern California,
the map created using complex eigenvectors has subtle differences with
the the map created using real-valued eigenvectors.  These differences
result in more useful information ({\it i.e.\/} a reduction in the map
entropy) and in better apparent correlation with future large
earthquakes.  Future monitoring and testing, however, will be
necessary to conclusively determine the accuracy and reliability of
complex PI analysis.


\section*{Acknowledgments}
This work has been supported by a grant from US Department of Energy,
Office of Basic Energy Sciences to the University of California, Davis
DE-FG03-95ER14499 (JRH and JBR) and through additional funding from
the National Aeronautics and Space Administration under grants through
the Jet Propulsion Laboratory to the University of California, Davis.



\end{document}